\documentstyle[epsf]{mn}
\newif\ifAMStwofonts    
\AMStwofontstrue

\ifoldfss
  \ifCUPmtlplainloaded \else
    \NewTextAlphabet{textbfit} {cmbxti10} {}
    \NewTextAlphabet{textbfss} {cmssbx10} {}
    \NewMathAlphabet{mathbfit} {cmbxti10} {} % for math mode
    \NewMathAlphabet{mathbfss} {cmssbx10} {} %  "   "    "
  \fi
  \ifAMStwofonts
    \ifCUPmtlplainloaded \else
      \NewSymbolFont{upmath} {eurm10}
      \NewSymbolFont{AMSa} {msam10}
      \NewMathSymbol{\upi}     {0}{upmath}{19}
      \NewMathSymbol{\umu}     {0}{upmath}{16}
      \NewMathSymbol{\upartial}{0}{upmath}{40}
      \NewMathSymbol{\leqslant}{3}{AMSa}{36}
      \NewMathSymbol{\geqslant}{3}{AMSa}{3E}

       \let\le=\leqslant
       
    \fi
  \fi
\fi % End of OFSS

\ifnfssone
  \newmathalphabet{\mathit}
  \addtoversion{normal}{\mathit}{cmr}{m}{it}
  \addtoversion{bold}{\mathit}{cmr}{bx}{it}
  \newmathalphabet{\mathbfit} % math mode version of \textbfit{..}
  \addtoversion{normal}{\mathbfit}{cmr}{bx}{it}
  \addtoversion{bold}{\mathbfit}{cmr}{bx}{it}
  \newmathalphabet{\mathbfss} % math mode version of \textbfss{..}
  \addtoversion{normal}{\mathbfss}{cmss}{bx}{n}
  \addtoversion{bold}{\mathbfss}{cmss}{bx}{n}
  \ifAMStwofonts
    \ifCUPmtlplainloaded \else
      \UseAMStwoboldmath
      \makeatletter
      \new@mathgroup\upmath@group
      \define@mathgroup\mv@normal\upmath@group{eur}{m}{n}
      \define@mathgroup\mv@bold\upmath@group{eur}{b}{n}
      \edef\UPM{\hexnumber\upmath@group}
      \new@mathgroup\amsa@group
      \define@mathgroup\mv@normal\amsa@group{msa}{m}{n}
      \define@mathgroup\mv@bold\amsa@group{msa}{m}{n}
      \edef\AMSa{\hexnumber\amsa@group}
      \makeatother
      \mathchardef\upi="0\UPM19
      \mathchardef\umu="0\UPM16
      \mathchardef\upartial="0\UPM40
      \mathchardef\leqslant="3\AMSa36
      \mathchardef\geqslant="3\AMSa3E

       \let\le=\leqslant

    \fi
  \fi
\fi % End of NFSS release 1

\ifnfsstwo
  \DeclareMathAlphabet{\mathbfit}{OT1}{cmr}{bx}{it}
  \SetMathAlphabet\mathbfit{bold}{OT1}{cmr}{bx}{it}
  \DeclareMathAlphabet{\mathbfss}{OT1}{cmss}{bx}{n}
  \SetMathAlphabet\mathbfss{bold}{OT1}{cmss}{bx}{n}
  \ifAMStwofonts
    \ifCUPmtlplainloaded \else
      \DeclareSymbolFont{UPM}{U}{eur}{m}{n}
      \SetSymbolFont{UPM}{bold}{U}{eur}{b}{n}
      \DeclareSymbolFont{AMSa}{U}{msa}{m}{n}
      \DeclareMathSymbol{\upi}{0}{UPM}{"19}
      \DeclareMathSymbol{\umu}{0}{UPM}{"16}
      \DeclareMathSymbol{\upartial}{0}{UPM}{"40}
      \DeclareMathSymbol{\leqslant}{3}{AMSa}{"36}
      \DeclareMathSymbol{\geqslant}{3}{AMSa}{"3E}

       \let\le=\leqslant

    \fi
  \fi
\fi % End of NFSS release 2

\ifCUPmtlplainloaded \else
  \ifAMStwofonts \else % If no AMS fonts
    \def\upi{\pi}
    \def\umu{\mu}
    \def\upartial{\partial}
  \fi
\fi

\title{Scaling of $N$-body calculations}
\author[H. Baumgardt]
       {H. Baumgardt \\
        Department of Mathematics and Statistics, University of Edinburgh, King's Buildings, Edinburgh EH9 3JZ}
\date{Accepted .
      Received ;
      in original form }

\pagerange{\pageref{firstpage}--\pageref{lastpage}}
\pubyear{2000}

\begin{document}

\maketitle

\label{firstpage}

\begin{abstract}
We report results of collisional $N$-body simulations aimed to study the
$N$-dependance of the dynamical evolution
of star clusters. Our clusters consist of equal-mass stars and are
in virial equilibrium. Clusters moving in external tidal fields and clusters
limited by a cut-off radius are simulated. Our main focus is to study the
dependence of the lifetimes of the clusters on the number of cluster stars and
the chosen escape condition.

We find that star clusters in external tidal fields exhibit a scaling problem
in the sense that their lifetimes do not scale with the relaxation time.
Isolated clusters show a similar problem if stars are removed only after
their distance to the cluster centre exceeds a certain cut-off radius. If 
stars are removed immediately after their energy exceeds the energy
necessary for escape, the scaling problem disappears. 
 
We show that some stars which gain the energy necessary for escape are
scattered to lower energies before they can leave the cluster.
Since the efficiency of this process decreases with increasing
particle number, it causes the lifetimes not to scale with the
relaxation time. Analytic formulae are derived for 
the scaling of the lifetimes in the different cases.  
 
\end{abstract}


\section{Introduction}
The aim of the present paper is to study the dependence of the
lifetimes of star clusters on the number of cluster stars and the chosen escape condition.
It is important to understand
this dependence, since at present it is impossible to perform a fully collisional simulation of 
globular clusters
with realistic particle numbers. Hence, one has to scale the results of 
simulations with smaller particle numbers to the globular cluster regime
(as for example in Wielen 1988), or adjust the parameters of other      
methods for star cluster evolution, for example Fokker-Planck calculations, such that 
they match the results of the largest feasible $N$-body calculations,
like in Takahashi \& Portegies Zwart \cite{tpz}. In both cases
it is important that the scaling of the lifetimes with the particle number is 
understood.

The theory for the dependence of the lifetime on the number
of cluster stars was developed by 
Ambartsumian \cite{am} and Spitzer \cite{s40}.
It is based on the assumption that the majority of more distant 
encounters between cluster stars is responsible for the mass-loss of the cluster. Distant
encounters tend to set up a Maxwellian velocity distribution at each point inside 
the cluster.
Such a distribution has non-zero density for every energy, so there are always stars 
which have velocities higher than the escape velocity of the cluster. These stars
will escape, 
causing a steady mass-loss of the cluster.  

Distant encounters lead to energy changes on the  
relaxation timescale (Chandrasekhar 1942, Spitzer 1987 eq.\ 2-62):     
\begin{equation}
t_{r} = \frac{0.065 \; v_m^3}{n \; m^2 \; G^2 \; ln \: \Lambda} \;\; ,
\end{equation} 
where $n$ is the density of cluster stars, $m$ the mean mass of a star, $v_m$ the 
average velocity of the stars, $G$ the constant of gravitation and $\Lambda$ is
proportional to the number of cluster stars.
During each relaxation time a constant fraction of cluster stars is scattered to energies
above the escape velocity, so the lifetimes of star clusters should be
multiples of their relaxation times.

There are however effects not accounted for by this theory.
H\'enon \cite{he60} for example studied isolated
clusters and showed that in this case the energy changes due to distant encounters are unimportant 
for escape, and instead most stars escape due to single   
close encounters with other cluster stars. In this and a later paper (H\'enon 1969), he
showed that this will lead to a scaling of the lifetime proportional to the number
of cluster stars times the crossing time of the cluster.

Another complication was first pointed out by Chandrasekhar \cite{ch} and studied in
detail by King \cite{k59}. Since stars with energies
high enough for escape need time to leave the cluster, some of them may be scattered
back to lower energies and become bound again. This reduces the number of stars
escaping from a cluster, thereby increasing its lifetime. If the escape time is constant, 
this effect will be more important for low-$N$  
clusters, since their relaxation times are shorter and a higher fraction of potential
escapers is retained. Backscattering therefore causes a deviation from a
scaling with the relaxation time.   

Further complications arise if external forces act upon a star cluster.
Clusters moving on elliptic orbits through their parent galaxies 
for example are subject to tidal heating, which acts on the orbital timescale and is 
independent
of the cluster's relaxation time. Since the changing tidal field removes stars, the 
lifetime of a cluster does not depend on the relaxation time alone.
Similar problems exist if star clusters have
to pass through galactic discs (Ostriker, Spitzer \& Chevalier 1972, Weinberg 1994ab) or 
the mass-loss of
the cluster stars is taken into account (Chernoff \& Weinberg 1990, Fukushige \& Heggie 1995).

Even for the simpler problem of a circular orbit with no individual mass loss of the cluster stars, 
the lifetime does not necessarily scale with the relaxation   
time. This was demonstrated by the Collaborative Experiment (Heggie et al.\ 1998), where
multi-mass clusters moving in circular orbits around a point-mass galaxy
were simulated.
Clusters containing between 1024 and 65536 stars were studied and it was found that
the lifetimes of the clusters increased more slowly than their relaxation times. Since
there is some uncertainty in the correct definition of the relaxation time for
a multi-mass cluster, it was however not clear if the observed discrepancy could not 
be removed by a different definition of the relaxation time.

It is the aim of the present paper to give
a better understanding of the dependence of the lifetime on the number of cluster stars. 
We begin by studying simpler clusters with a tidal cut-off
and use the results obtained there to understand the
behaviour of clusters in a steady tidal field.

\section{Description of the runs}

The calculations were performed with the collisional Aarseth $N$-body code NBODY6++
(Makino \& Aarseth 1992, Aarseth 1999). This code uses an Hermite integration scheme with 
block time-steps and Ahmad-Cohen neighbour scheme for the integration. 
It has recently been parallelised (Spurzem 1999, Spurzem \& Baumgardt 2000) by means of 
MPI-routines to increase its peak performance.   

All our clusters consist of equal-mass stars and their density distributions
are given by $W_0 = 3.0$ King profiles. The
tidal radii of the King models are adjusted such that they are equal to the
cut-off radii for the isolated clusters, and are equal to the tidal radii given by the
galactic tidal fields in the models with a full tide.
 
Clusters containing between 128 and 16384 stars
were simulated. Small $N$ clusters were simulated more than once
in order to reduce the statistical noise. The evolution of  
small $N$ clusters was followed on a Pentium III PC, while
clusters with $N=16384$ stars were simulated on a CRAY T3E parallel
computer using 8 or 16 processors. With 8 processors, it took about 550 CPU-hours 
to follow the evolution of a King $W_0 = 3.0$ cluster with 16384 stars 
until complete dissolution. 
 
Three different types of runs were performed: First we studied isolated clusters
and removed stars if their energies were large enough so that they could 
reach the tidal radius. 
In the second case we also studied isolated clusters, but removed stars if 
their distance to the cluster centre exceeded the tidal radius. Simulations of this kind 
are often used to study tidally limited clusters. We finally studied
clusters moving on circular orbits around point-mass galaxies with a proper tidal field.

Table 1 gives an overview of the simulations performed. The columns contain from left to right the
number of cluster stars $N$, the number $N_{Sim}$ of simulations performed, the mean time required to lose
half the mass and an error estimate for the half-mass time. The error estimate was derived by 
calculating the standard deviation of the individual runs around the mean half-mass time and dividing 
it by the square root of $N_{Sim}$. We use the half-mass time to study the scaling
in order to avoid very low $N$ effects. These might play a role for the smallest clusters 
at the end of their lifetime.

All times are given in $N$-body units, where the total mass and energy of a cluster
are given by $M=1$ and $E_C~=~-0.25$ initially, and the constant of gravitation $G$ is equal to~1.
We will use these units throughout the paper.

\begin{table}
\caption[]{Details of the performed $N$-body runs.}
\begin{tabular}{rrr@{.}lr}
\noalign{\smallskip}
 & \multicolumn{4}{c}{Energy Cutoff}\\[+0.2cm]
\multicolumn{1}{c}{$N$}& \multicolumn{1}{c}{$N_{Sim}$} & \multicolumn{2}{c}{$T_{Half}$} & \multicolumn
{1}{c}{$\sigma_{T_H}$} \\
\noalign{\smallskip}
  128 & 128 &    53 & 5 &  1.0 \\
  256 &  96 &    85 & 6 &  1.0 \\
  512 &  32 &   141 & 6 &  1.3 \\
 1024 &  8  &   239 & 5 &  4.8 \\
 2048 &  4  &   423 & 6 &  6.4 \\
 4096 &  2  &   739 & 6 & 16.0 \\
 8192 &  2  &  1320 & 9 & 33.0 \\
16384 &  1  &  2371 & 5 & \\
\end{tabular}
\vspace*{0.3cm}  
\begin{tabular}{rrr@{.}lr}
\noalign{\smallskip}
 & \multicolumn{4}{c}{Radial Cutoff}\\[+0.2cm]
\multicolumn{1}{c}{$N$}& \multicolumn{1}{c}{$N_{Sim}$} & \multicolumn{2}{c}{$T_{Half}$} & \multicolumn
{1}{c}{$\sigma_{T_H}$} \\
\noalign{\smallskip}
  128 & 96  &  107 & 3 &  2.3 \\
  256 & 64  &  155 & 8 &  1.9 \\
  512 & 32  &  234 & 6 &  2.5 \\
 1024 & 16  &  363 & 0 &  3.2 \\
 2048 &  8  &  585 & 0 &  9.1 \\
 4096 &  4  &  996 & 8 & 10.8 \\
 8192 &  2  & 1704 & 5 &  7.3 \\ 
16384 &  1  & 2894 & 5 & \\
\end{tabular}
\vspace*{0.3cm}
\begin{tabular}{rrr@{.}lr}
\noalign{\smallskip}
 & \multicolumn{4}{c}{Full tidal field}\\[+0.2cm]
\multicolumn{1}{c}{$N$}& \multicolumn{1}{c}{$N_{Sim}$} & \multicolumn{2}{c}{$T_{Half}$} & \multicolumn{1}{c}{$\sigma_{T_H}$} \\
\noalign{\smallskip}
  128 & 128 &    89 & 4 &   1.2  \\ 
  256 &  64 &   126 & 7 &   1.6  \\
  512 &  32 &   182 & 6 &   3.8  \\  
 1024 &  16 &   258 & 5 &   4.2  \\
 2048 &   8 &   372 & 9 &   6.6  \\  
 4096 &   4 &   558 & 1 &   5.2  \\ 
 8192 &   2 &   840 & 9 &   1.1  \\ 
16384 &   1 &  1176 & 8 &
\end{tabular}
\end{table}

\section{Results}
 
\subsection{Energy cut-off models}

\begin{figure*}
\epsfxsize=17cm
\epsffile{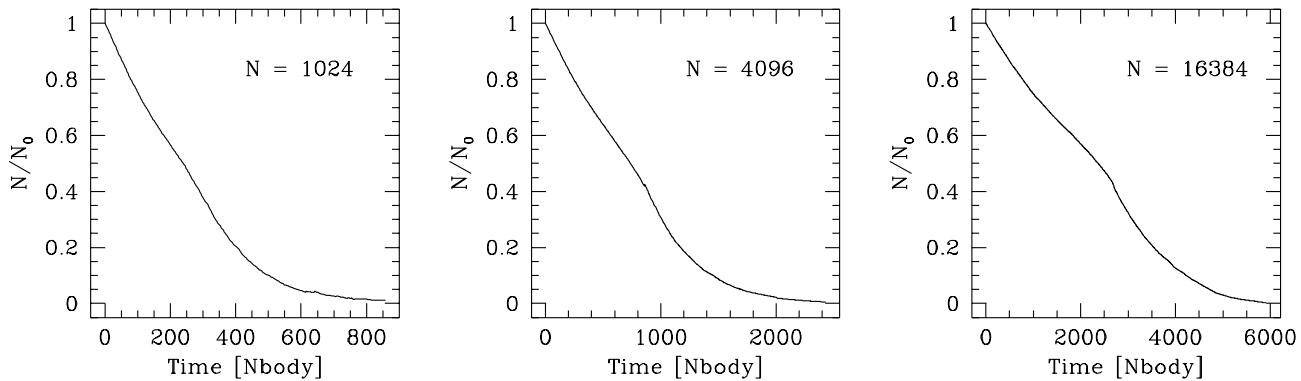}
  \caption{Evolution of the fraction of bound stars as 
    a function of $N$-body time for three energy cut-off clusters. Shown
  is the mean of all clusters containing 1024, 4096 and 16384 stars initially.}
\end{figure*}
 
In the energy cut-off case, we immediately remove stars once their energies 
become high enough so that they can reach the tidal radius $r_t$ of the cluster.
The maximum distance $r_m$ that a star at a distance $r$ from the cluster centre can
reach, provided it does not experience any encounters with other cluster stars, is
given by the following equation:
\begin{equation}
\phi(r_m) + \frac{1}{2} v^2_\perp (\frac{r}{r_m})^2 = \phi(r) + \frac{1}{2} ( v^2_{\vert|}
+ v^2_\perp )
\end{equation}
where $\phi(r)$ is the potential at position $r$ and $v_{\vert|}$ and $v_\perp$
denote the velocity components parallel and perpendicular to the direction 
from the star's position to the   
cluster centre. Note that the second term on the left hand side has to be
added due to the conservation of angular momentum. Since we remove stars if they
can reach the tidal radius, the energy $E_{Crit}$ necessary for escape is given by:
\begin{equation}
E_{Crit} = - \frac{M_C}{r_t} + 0.5 \cdot \frac{\vec{L}^2}{r_t^2}  \;\; ,
\end{equation}
where $L$ denotes the angular momentum of the star with respect to the cluster centre 
and $M_C$ is the present mass of all
stars still bound to the cluster. We check the energy of each star while it is advanced
in the regular integrational part of NBODY6++ and all stars with energies larger 
than their critical energy $E_{Crit}$ are immediately removed. The tidal radius $r_t$ is 
kept fixed during the calculation in order to minimize the influence of drift in
energy space of individual cluster stars due to the mass-loss of the cluster.
Our model resembles many Fokker-Planck or gaseous 
models for star cluster evolution, in which the tidal field is treated as an energy boundary, 
and stars beyond this boundary are immediately removed.  

Figure 1 shows the evolution of the number of bound stars (for the energy cut-off models equal to 
all stars still in the simulation) 
% as a function of time  
for three energy cut-off clusters. The number of bound stars decreases almost linearly until 90\%
of them are lost. There is a slight increase in the mass-loss rate at around core-collapse
(which occurs after 60\% of the stars are lost, see Fig.\ 3). The slow-down of the 
mass-loss at the end of the simulations can be explained by the constant tidal radius of 
our clusters. Due to this the outer lagrangian radii also remain nearly constant, so the 
crossing time becomes very large at the end. Hence the clusters evolve
only slowly. It is therefore better to use the half-mass times of the  
clusters to study the scaling. 
This also avoids effects due to the core-collapse of the clusters.
\begin{figure}
\begin{center}
\epsfxsize=8.0cm
\epsffile{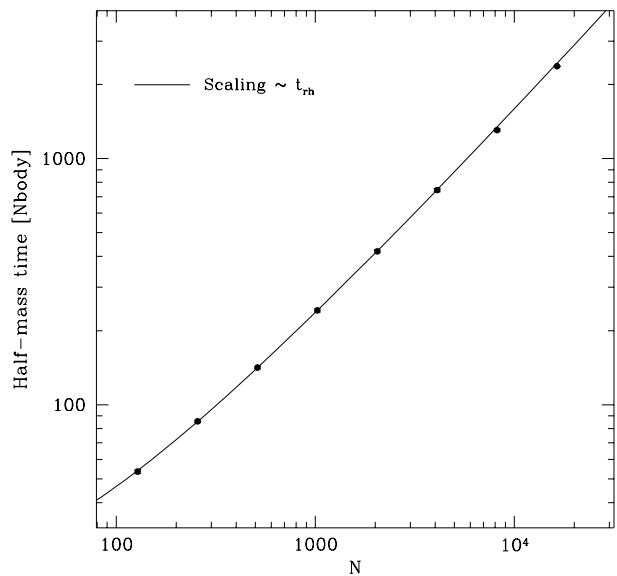}
\end{center}
  \caption{Mean half-mass times as a function of the number of cluster stars
  for the energy cut-off clusters. The solid line shows a theoretical
  scaling with the relaxation time, fitted to the mean half-mass time for $N=1024$.
  It provides an excellent fit to the half-mass times of the $N$-body runs
  (filled circles).}
\end{figure}

Figure 2 shows the half-mass times as a function of the number of cluster stars.
The solid line shows a scaling proportional to the half-mass relaxation time, fitted
to the results of the $N=1024$ runs. The half-mass relaxation time was taken
from Spitzer (1987), eq.\ 2-63                           
\begin{equation}
t_{rh} = 0.138 \frac{\sqrt{N} \; r_h^{3/2}}{\sqrt{m} \; \sqrt{G} \; \ln(\gamma N)} \;\; ,
\end{equation}
with the value of the Coulomb logarithm  
taken to be $\gamma = 0.11$. Such a value was obtained by Giersz \& Heggie \cite{gh} by a 
comparison of the evolution of single-mass clusters containing $N=500$ and $N=2000$ stars
respectively. As can be seen, a scaling proportional to the half-mass relaxation time provides 
an excellent fit to the $N$-body results.     
\begin{figure}
\begin{center}   
\epsfxsize=8.0cm
\epsffile{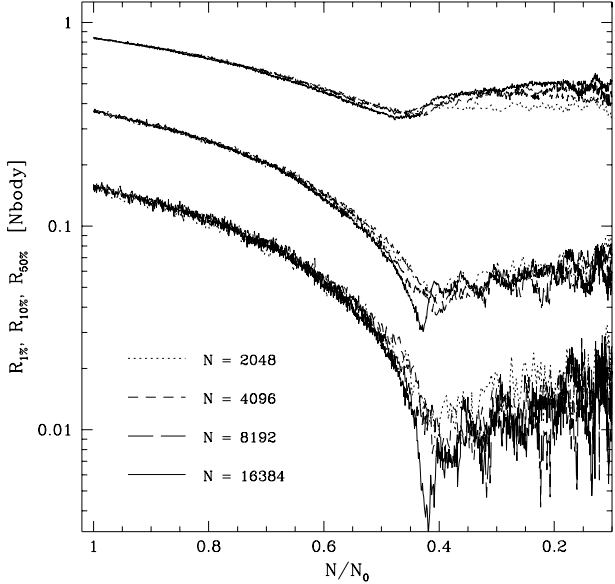}
\end{center} 
 \caption{Evolution of the lagrangian radii containing 1, 10 and 50 \% of all stars,
  as a function of the number of stars for clusters with $2048 \le N \le 16384$ initially. 
   The graphs for different $N$ lie on top of each other, showing that both quantities 
    change on the same timescale.}   
\end{figure}
Following an idea of Toshio Fukushige, we can also check the scaling of the lifetimes without 
adopting a specific formula
for the relaxation time. Figure 3 shows the evolution of the lagrangian radii
as a function of the number of stars still bound to the clusters for clusters
containing between $2048 < N < 16384$ stars initially.  
The change in the lagrangian radii is due to the core-collapse,
which is generally believed to happen on the relaxation timescale since it is
driven by two-body relaxation. If the mass-loss of the clusters is also happening on the
relaxation timescale, one expects the graphs for different $N$
to lie on top of each other. Otherwise differences in the scaling laws of both
quantities should create different curves for different $N$.
As can be seen in Fig.\ 3, the curves lie on top of each other, at least in the pre-collapse phase.
The differences between the clusters around and after core-collapse are due to
the different degrees of central concentration the clusters reach for different initial
particle number and are not in contradiction with the pre-collapse evolution.      

Combining the results of Figures 2 and 3, we conclude that the dissolution of the
energy cut-off clusters happens on the relaxation timescale, and that eq.\ 4 with 
$\gamma = 0.11$ provides a good description for the relaxation time of single-mass clusters.
We will therefore use it to study the scaling of the other models.

\subsection{Radial cut-off models} 

The radial cut-off case differs from the models in the previous section only by the escape
condition. Here, we remove stars after they have crossed the tidal radius $r_t$ of the
cluster, i.e.\ their distance to the cluster centre exceeds $r_t$.
As in the energy cut-off case, the tidal radius is kept fixed during the integration. 
 
Figure 4 shows the half-mass times of the radial cut-off clusters. The solid line shows
a scaling with the relaxation time, fitted to the result of the largest run.
There is a clear deviation from such a scaling. The half-mass times are sufficiently close 
to the expected curve only for the two largest models. Otherwise, 
they increase more slowly with $N$ than the relaxation time.
% (note that errorbars are plotted in Fig.\ 3).    

In the radial cut-off models stars need time to travel from the place where they are
scattered above the critical energy to the tidal radius of the cluster.
While they move outward, potential escapers may be scattered back to
lower energies and become bound again.
This decreases the number of stars escaping from a cluster within a certain interval
of time. To study the influence of backscattering, we divide the stars  
into three categories (bound stars with $E<E_{Crit}$, potential escapers with energies $E>E_{Crit}$,
and escaped stars) and consider the processes shown in
Figure 5: Stars are scattered into and out of the potential escaper regime on relaxation
timescales and leave the clusters within one escape time. All three processes can be
expected to be in equilibrium with each other, since the escape times are much shorter than
the lifetimes of the clusters (see Table 2).
We therefore obtain for $N_{PE}$:
\begin{equation}
\frac{d N_{PE}}{dt} = k_1 \frac{1}{t_{rh}} N_{Bound} - k_2 \frac{1}{t_{rh}} N_{PE} - \frac{1}{t_e} N_{PE} = 0
\end{equation}
with the solution
\begin{equation}
N_{PE} = N_{Bound} \; \frac{k_1 t_e}{t_{rh} + k_2 t_e} \;\; .
\end{equation}
Here  $N_{Bound}$ is the number of all stars with energies $E<E_{Crit}$ and $k_1$ and $k_2$
are constants which reflect the efficiencies for scattering stars above and below the
critical energy. If the cluster mass decreases linearly with time,
the lifetimes of the clusters (or in our case the half-mass times) can be estimated by 
dividing the number of all stars $N_{Star}$ in the cluster by 
the number of stars escaping from the cluster within a given time interval. Hence
\begin{eqnarray}
T_{Half} & = & \frac{\frac{1}{2} N_{Star}}{\frac{1}{t_e} N_{PE}} \\ \nonumber
 & = & \frac{1}{2 k_1} ( t_{rh} + (k_1 + k_2) t_e) \;\;.
\end{eqnarray}
This solution has two regimes. If $t_{rh} \gg (k_1 + k_2) t_e$, backscattering is unimportant
since the timescale for it is much larger than the escape time. 
Hence, all stars scattered above the critical energy will escape, and the lifetime
scales with the relaxation time.  
For smaller $t_{rh}$, backscattering leads to an increase of the lifetimes.

In order to fit our results, we have to determine the unknown quantities $k_1$, 
$k_2$ and $t_e$.
The escape time can be measured in the simulations: For each escaping star we take 
the difference between the time it leaves the cluster and its last upward crossing of the 
critical energy $E_{Crit}$. This is done until the half-mass time is reached for 
a particular simulation and the mean over all simulations is taken.
Table~2 gives the mean escape times determined that way. $t_e$ increases slightly with
$N$ since potential escapers in high-$N$ clusters acquire less 
energy before they leave the cluster due to the longer relaxation time. It therefore takes 
more time until they reach the tidal radius. In addition, 
the fraction of stars that escape due to large-angle encounters, which have large
velocities and correspondingly small escape times, may drop with increasing $N$.

The constants $k_1$ and $k_2$ are best determined from a fit to the data.
We find that $k_1 = 0.053$ and $k_2 = 1.01$ give the best fit. Figure 4 compares the 
predicted lifetimes with the $N$-body data for this choice of constants. There 
is good agreement between both, so a model with backscattering explains the 
$N$-dependence of the lifetimes.    

The value required for $k_1$ means that high-$N$ clusters lose a fraction 
$k_1 = 5.3  \cdot  10^{-2}$ of their mass during each relaxation time. This is only slightly
higher then the value found by Spitzer (1987, eq.\ 3-27) $\xi_e = 4.5 \cdot 10^{-2}$ for
the evolution of H\'enon's self-similar model. It is also close to the values
found by Johnstone \cite{jo} from Fokker-Planck simulations of single-mass
clusters surrounded by a tidal cut-off. Since he did not study King models with a
central concentration of $W_0 = 3.0$, no direct comparison is possible, but judging
from his results for $W_0 = 2.0$ and $W_0 = 4.0$, it seems that our mass-loss rate
is again slightly larger. The reason may be that the Fokker-Planck approach neglects
close encounters, which may be important in the cores of the clusters
and contribute to the mass-loss.   

The value for $k_2$ is rather high, since it implies that the process of backscattering
is some 20 times more effective than the scattering of stars above the critical energy.
It can be explained by the fact that stars are drifting only slowly through energy space,
so a typical potential escaper has an energy only slightly above the critical energy.
% If $t_{r}$ is the relaxation time of a potential escaper, it will travel a distance
% $\Delta E = \sqrt{t_{esc}/t_{r}}$ beyond $E_{Crit}$ before it escapes. To be scattered from such an 
% energy to energies below $E_{Crit}$ also takes one relaxation time $t_r$. $t_r$ should not be very much
% different from the half-mass relaxation time $t_{rh}$, hence $k_2$ must be of order unity. 
It is therefore easily scattered back to lower energies and becomes bound again, whereas it
is much harder to scatter a bound star to energies above $E_{Crit}$.
\begin{figure}
\epsfxsize=8.0cm
\begin{center}
\epsffile{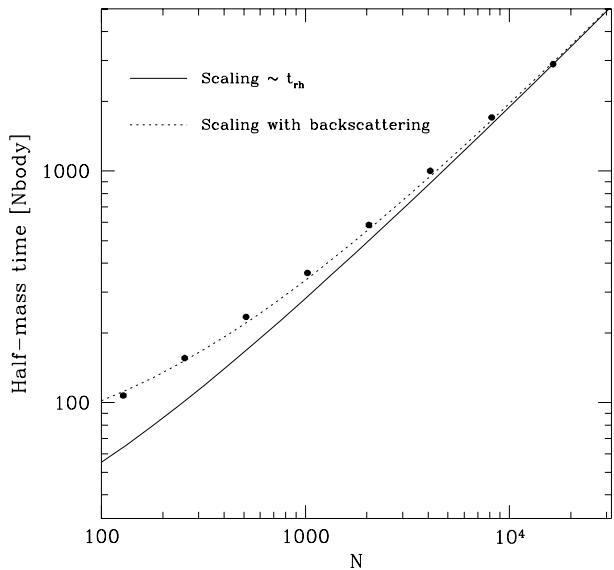}
\end{center}
  \caption{Mean half-mass times as a function of the number of cluster stars for the radial
   cut-off models. The solid line shows a scaling with the relaxation time fitted to
     the result of the largest run. There is a clear deviation from such a scaling.
        The dashed line shows the fit obtained by taking the influence
         of backscattering into account.}
\end{figure}

\begin{table}
\caption[]{Mean escape times $t_e$ and potential escaper fraction 
$F_{PE}$ for the radial cut-off models. Shown is the mean over all
simulations calculated up to the half-mass time.}      
\begin{center}
\begin{tabular}{rcc}
\noalign{\smallskip}
\smallskip
  $N$ & $t_e$  &  $<F_{PE}>$ [\%]  \\
  128 &  4.71  &   4.41   \\    
  256 &  4.76  &   2.97  \\
  512 &  4.99  &   2.00   \\ 
 1024 &  5.23  &   1.29   \\
 2048 &  5.34  &   0.82   \\
 4096 &  5.55  &   0.49   \\
 8192 &  5.95  &   0.29   \\
16384 &  6.43  &   0.14   
\end{tabular}
\end{center}
\end{table}

We finally compare the number of potential escapers in the $N$-body runs with our 
prediction. Table 2 lists the mean fraction of potential escapers, defined
as $F_{PE} = \frac{N_{PE}}{N_{Star}}$, calculated from the beginning up to the 
half-mass times of the clusters. It is compared with eq.\ 6, with the relaxation
time calculated at the point when the clusters have lost 25~\% of their stars:
\begin{equation}
<F_{PE}> \; = \; <\frac{N_{PE}}{N_{Star}}>\; \; \approx \;  
 \frac{k_1 t_e}{(k_1+k_2) t_e + t_{rh (0.75 N_0)}} \;\;.
\end{equation}
Figure 6 compares the number of potential escapers with the predicted fraction.
They both decrease with increasing $N_0$ due to the increase
in the relaxation times and the predicted fraction gives a very good fit to the 
$N$-body results.

We conclude that the lifetimes of the radial cut-off models are influenced
by backscattering. This process increases the lifetimes of low $N$-clusters.
We expect that backscattering becomes unimportant for large enough $N$, 
since the relaxation time increases until all stars scattered above the
critical energy will escape.   
Similar results were also found by King \cite{k59}. The main difference 
between his results and our work is that due to the small excess energies of potential
escapers, backscattering is more important
for radial cut-off clusters than estimated by him.
\begin{figure}
\epsfxsize=5.0cm
\begin{center}
\epsffile{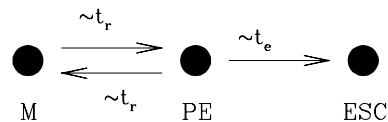}
\end{center}
  \caption{Model for the evolution of the radial cut-off clusters. Bound members (M) are scattered
  above the critical energy required for escape and become potential escapers (PE).
  Potential escapers are either scattered back before they can leave the cluster and become bound
  members again, or escape.}
\end{figure}

\begin{figure}
\epsfxsize=8.0cm
\begin{center}
\epsffile{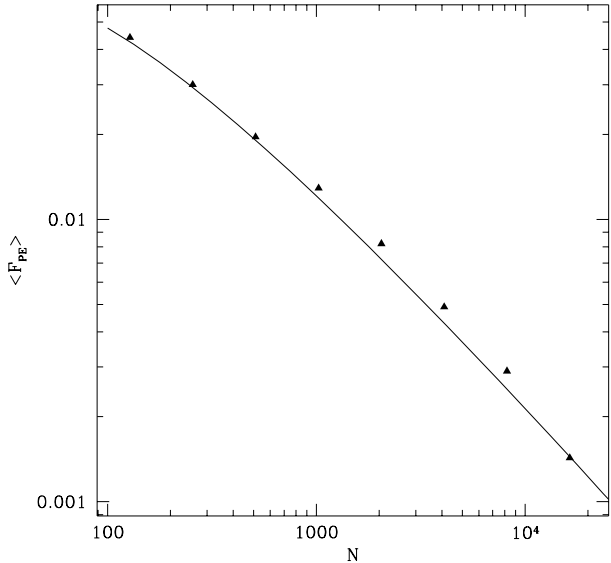}
\end{center}
  \caption{Mean potential escaper fraction as a function of the number of cluster stars. Triangles
   show the $N$-body results. The solid line shows the behaviour predicted by our model for 
    backscattering.}
\end{figure}
 
\subsection{Clusters in a steady tidal field}

We finally discuss the evolution of clusters moving in circular orbits around point-mass
galaxies. In these models, the full tidal field is taken into account and stars are removed if
their distance to the cluster centre exceeds twice the tidal radius. We note that the removal 
of stars has no influence on the scaling since it is made at a radius where nearly all
stars are already unbound to the clusters. 

In a constant tidal field, the tidal radius $r_t$ and the critical energy required for
escape are given by
\begin{eqnarray}
r_t = \sqrt[3]{\frac{M_C}{3 M_G}} \; R_G\\
E_{Crit} = -\frac{3}{2} \frac{M_C}{r_t} \; ,
\end{eqnarray}
where $R_G$ is the radius of the cluster orbit and $M_G$ the mass of the galaxy. 
Since the critical energy gives
only a necessary but not a sufficient criterion for escape,
some stars can remain trapped inside the potential well even if their energies
exceed $E_{Crit}$. For the other stars with $E>E_{Crit}$, the problem
of their escape time was studied by Fukushige \& Heggie
\cite{fh}. They found that the time required for escape
from a fixed potential is mainly a function of the excess energy $\Delta E = (E-E_{Crit})$
and drops approximately with this energy difference to the second power:
\begin{equation}
t_{e} \propto \left(\frac{E_{Crit}}{E-E_{Crit}}\right)^2 \; .
\end{equation}
This dependence arises since stars with energies slightly above the critical one can escape
only through small apertures around the lagrangian points $L_1$ and $L_2$,
which lie along the line connecting the cluster centre and the galaxy. These apertures become
smaller and smaller as $E$ approaches $E_{Crit}$. Hence
stars have to pass through the cluster an increasing number of times before they find a hole in the
potential well. The mean time required for escape is therefore much higher
than in the radial cut-off case and backscattering of potential escapers should happen
more often. In addition, potential escapers will also drift to higher energies because the
cluster loses mass while they are still trapped inside the potential well of the cluster. This effect
will certainly influence the number of potential escapers and shorten the lifetimes of the clusters. 
However, since it is happening on the dissolution timescale, it does not influence the scaling law.
\begin{figure}
\epsfxsize=8.0cm
\begin{center}
\epsffile{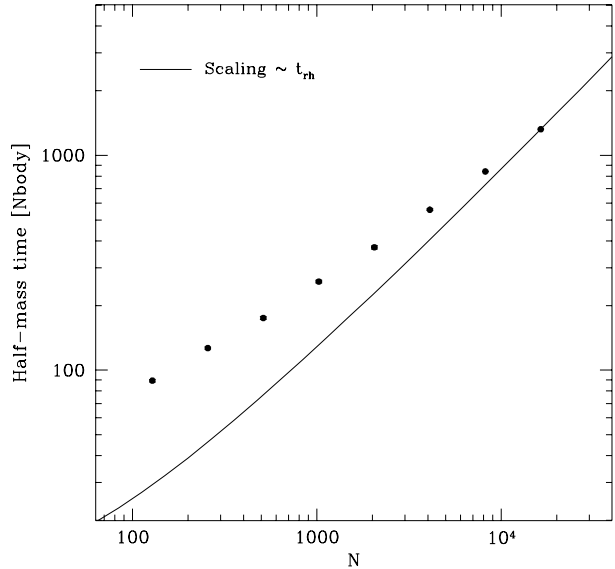}
\end{center}
\caption{Same as Figure 2, but now for clusters in an external tidal field. 
  There is a clear deviation from the expected scaling with the relaxation time. This deviation
   remains up to the highest simulated $N$. Compared to the highest run, the mean half-mass 
    time of the $N=128$ runs is a factor of 3 larger than expected.}
\end{figure}
\begin{figure}
\epsfxsize=8.0cm   
\begin{center}
\epsffile{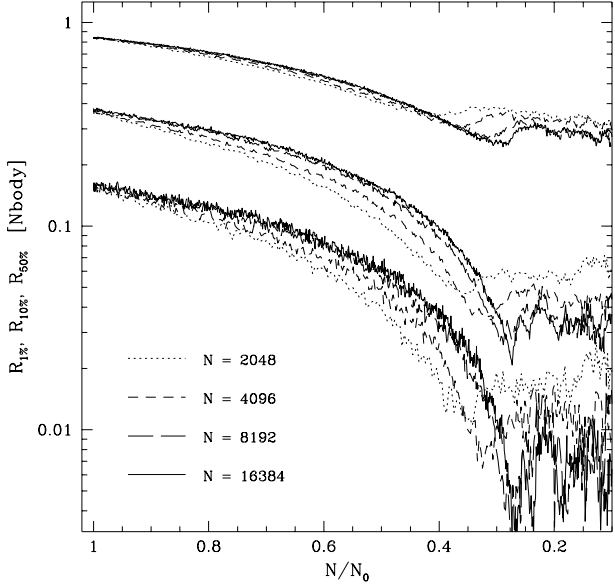}
\end{center} 
 \caption{Same as Figure 3, but now for clusters in an external tidal field. The curves for 
   different $N$ do not lie on top of each other, instead core-collapse happens later for
    clusters with higher initial particle numbers. This confirms that the mass-loss is not
     happening on the relaxation timescale, in agreement with Figure~7.}  
\end{figure}

Figure 7 shows the half-mass times as a function of the initial number of cluster stars.
Compared to the radial cut-off clusters the discrepancy between a scaling with
the relaxation time and the $N$-body results is larger and there is no sign that this discrepancy
vanishes for higher particle numbers. One reason for the larger discrepancy between theory and 
$N$-body results is certainly the longer time that is required for escape in a tidal field. 
Figure 8 shows the evolution of the lagrangian radii as a
function of the number of bound stars. The curves for different
$N$ do not lie on top of each other, instead core-collapse happens later for clusters
with higher initial particle numbers. This means that the timescale for mass-loss differs 
from the timescale for core collapse, in agreement with Fig.~7. Summarising, Figs.\ 7 and~8
indicate that the lifetime does not scale with the relaxation time.

In order to understand the results of the $N$-body runs, we will neglect
the fact that there are stars with $E > E_{Crit}$ that can never escape, and also the energy change of 
the
stars due to the mass-loss of the cluster. We will also neglect dynamical friction, its influence will be
discussed later. We take the energy dependence of the escape  
times into account, since the mean energy of potential escapers will change as a function of the 
number of cluster stars. Our model is comparatively simple, but should give an approximation
to the processes happening in the $N$-body simulations.

We split the potential escaper regime into different energy 
bins $E$ with particle numbers $N(E)$ (see Fig.\ 9). Utilising 
the expression for the escape times found by Fukushige \& Heggie (eq.\ 11 of our paper), 
we obtain for the change of $N(E)$ with time:
\begin{equation}
\frac{d N(\hat{E})}{d t} = \frac{k_1}{t_{rh}} \frac{d^2 N(\hat{E})}{d \hat{E}^2}-\hat{E}^2 \; \frac{N(\hat{E})}{t_{esc}}
\end{equation}
Here the variable $\hat{E} = (E-E_{Crit})/E_{Crit}$ was introduced and $t_{esc}$ is the
time required for escape at energy $\hat{E}~=~1$.
Requiring equilibrium $d N(\hat{E})/dt = 0$ gives the following solution for $N(\hat{E})$:   
\begin{equation}    
N(\hat{E}) \propto \sqrt{\hat{E}} \; K_{\frac{1}{4}}\left(\frac{1}{2} \sqrt{t_{rh}/(k_1 \, t_{esc})} \hat{E}^2\right)
\end{equation}    
with $K_{1/4}$ being a modified Bessel-function. Escape takes infinitely long for a star
with zero excess energy, so the number of stars at $\hat{E} = 0$ is solely determined by the 
scattering of stars into the potential escaper regime and the backscattering of potential 
escapers, and should be proportional to 
the number $N_{Star}$ of cluster stars: 
\begin{equation}
N(\hat{E}) \propto N_{Star} \, \left( \frac{t_{rh}}{t_{esc}} \right)^{1/8} \, \sqrt{\hat{E}} 
\; K_{\frac{1}{4}}\left(\frac{1}{2} \sqrt{t_{rh}/(k_1 t_{esc})} \hat{E}^2\right)
\end{equation}
The mass-loss rate is given by
\begin{eqnarray}   
\dot{N}_{Esc} & = & \frac{1}{t_{esc}} \int \hat{E}^2 \, N(\hat{E}) \, d\hat{E} \\ \nonumber
         & \propto & N_{Star} \; \; t_{rh}^{-3/4} \; \;  t_{esc}^{-1/4}
\end{eqnarray}    
and dividing the number of bound stars by $\dot{N}_{Esc}$ gives the following relation for the 
life time $t_h$:
\begin{equation}    
t_h \propto t_{rh}^{3/4} \;  t_{esc}^{1/4} 
\end{equation}
Hence, although the energy changes inside the cluster are assumed to happen on the relaxation 
time scale, we obtain the rather surprising result that the dissolution time
scales with the relaxation time to the power of 3/4.      
\begin{figure}
\epsfxsize=8.3cm
\begin{center}
\epsffile{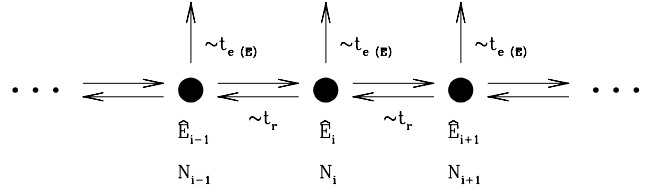}
\end{center}
\caption{Model for the evolution of clusters in a steady tidal field. The potential 
  escaper regime is split into different energies $\hat{E}$. Stars change their energies
  on the relaxation timescale and leave the cluster during the escape time $t_e$.  
  The escape time drops with the energy difference $\hat{E}=(E-E_{Crit})/E_{Crit}$ to the second 
  power $t_e \propto \hat{E}^{-2}$.}  
\end{figure}
\begin{figure}
\epsfxsize=8.0cm
\begin{center}
\epsffile{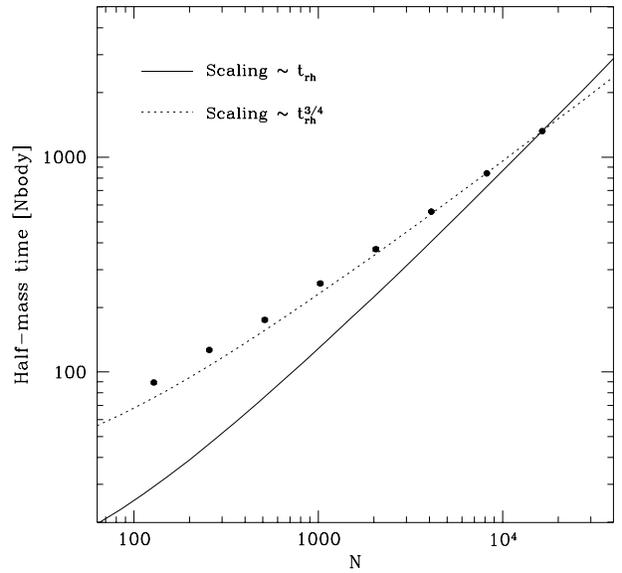}
\end{center}
\caption{Comparison of the predicted $t_{rh}^{3/4}$ scaling with the $N$-body data for clusters
    in a
     tidal field. The theoretical curves are adjusted such to match the result of the highest run.
    A scaling proportional to $t^{3/4}_{rh}$ gives a good fit to the results of the
   $N$-body runs and is predicted by our theory.}
\end{figure}

Figure 10 compares the $t^{3/4}_{rh}$ scaling with the $N$-body results. The agreement is 
good, the half-mass times in the $N\mbox{-body}$ models increase only slightly slower with $N$ than 
predicted. The reason for the small discrepancy may be that our clusters don't start with 
a potential escaper distribution that is in equilibrium. Our clusters start with primordial escapers
since they are set up such that no star crosses the tidal radius only 
if the clusters are isolated. Since the tidal field adds a force which alters the potential 
energy of the cluster stars, some stars will initially have energies $E > E_{Crit}$. 
Their number and energy distribution will certainly not be the 
equilibrium one, so the initial phases until an equilibrium distribution is reached
will show a different scaling. This may explain the slight differences.   

Figure 11 shows the evolution of the potential escaper fraction with time. All   
clusters contain 15 \% potential escapers initially due to the set-up. The slight increase
of $F_{PE}$ in the low-$N$ clusters probably indicates the phase where the clusters
evolve towards equilibrium. After equilibrium is reached, which happens at about $N/N_0 = 0.9$,
the fraction of potential escapers
\begin{figure}
\epsfxsize=8.0cm
\begin{center}
\epsffile{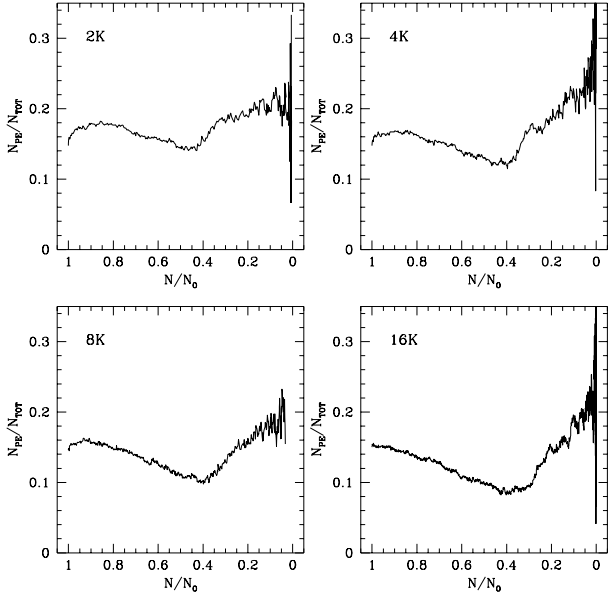}
\end{center}
\caption{Potential escaper fraction for clusters in a tidal field as a function of the fraction of
   stars still bound to the clusters. The fraction drops until core-collapse and rises afterwards
   due to the increase of potential escapers from the core.} 
\end{figure}
drops until core-collapse. Core-collapse then causes a sharp increase in
the fraction of potential escapers, and all models end up with a potential escaper
fraction of about 20~\%. 

If we integrate the solution for $N(\hat{E})$ over all energies, our theory gives the 
following result for the dependance of the potential escaper fraction on the initial
number of cluster stars:
\begin{eqnarray}
F_{PE} & = \frac{N_{PE}}{N_{Star}} & = \frac{1}{N_{Star}} \int N(\hat{E}) d\hat{E}\\ \nonumber
       & \propto t_{rh}^{-1/4} &
\end{eqnarray}  
Figure 12 compares the prediction with the mean escaper fraction of the $N$-body runs. 
To avoid effects due to the initial evolution, the mean fraction in the $N$-body runs
was calculated between the time the clusters lost 10~\% of their
mass and the half-mass time. Both fractions decrease and we obtain a fit to the $N$-body
results for clusters with $N \le 1024$. Later the potential escaper fraction drops less quickly 
in the $N$-body results than in our theory. Part of this discrepancy is certainly due to bound 
members
that have $E>E_{Crit}$ and that were neglected in our theory. The slow decrease means that 
even in clusters with particle numbers 
as high as globular clusters, several percent of the stars will have energies above the 
critical one.
\begin{figure}
\epsfxsize=8.0cm
\begin{center}
\epsffile{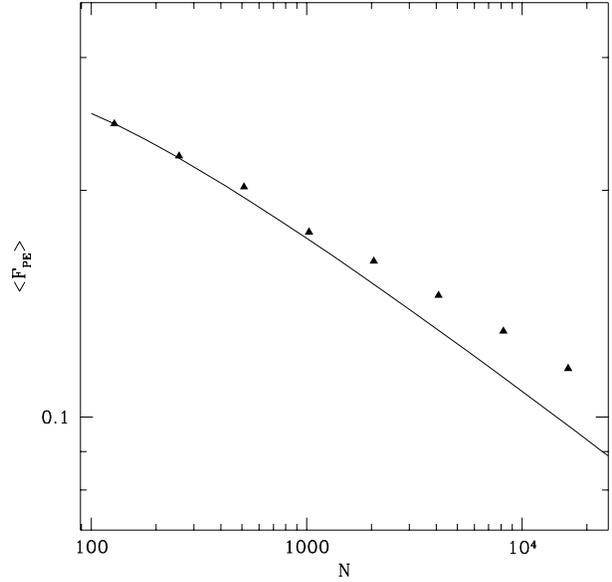}
\end{center}
\caption{Mean potential escaper fraction as a function of the number of cluster stars. The solid 
line shows our prediction fitted to the results of the smallest runs. It provides a good fit only 
for $N \le 1024$. Afterwards it drops too quickly, one reason being that bound members with 
$E>E_{Crit}$ were not taken into account.}     
\end{figure}

Our results for the scaling of the lifetimes do not change if we add an energy drift term due to the 
mass-loss of the
clusters to the right side of eq.\ 12. This might be expected, since the drift in energies is happening
on the mass-loss timescale itself. If we add a term which is due to dynamical friction, eq.\ 12 
becomes
\begin{equation}
\frac{d N(\hat{E})}{d t} = \frac{k_1}{t_{rh}} \frac{d^2 N(\hat{E})}{d \hat{E}^2} + \frac{k_2}{t_{rh}} 
\frac{d N(\hat{E})}{d \hat{E}} - \hat{E}^2 \; \frac{N(\hat{E})}{t_{esc}} \;\; .      
\end{equation}
The corresponding solution for $N(\hat{E})$ are Whittaker functions. Numerical exploitation of the solution shows that
if dynamical friction tends to slow down potential escapers, i.e. $k_2 > 0$, the scaling becomes flatter than
in the case without friction. The change in the scaling vanishes for large $N$, in which case the dissolution times
always scale proportional to $t_{rh}^{0.75}$.

The results presented so far were obtained for single-mass clusters. Figure 13 compares the 
predicted scaling with the half-mass times of multi-mass clusters.    
The data was taken from runs made by Sverre Aarseth and Douglas Heggie for 
the Collaborative Experiment. Their clusters had a Salpeter like mass-function, but are otherwise 
identical to the clusters studied here. In order to calculate the relaxation time, a value of 
$\gamma = 0.02$ was assumed for the Coulomb logarithm (Giersz \& Heggie 1996).

We obtain a fairly good agreement with our prediction since the half-mass times scale only slightly 
steeper than with $t_{rh}^{3/4}$. The slight difference may be due to the core-collapse of the 
clusters. For the multi-mass clusters core-collapse happens before half-mass is reached and
it happens earlier for smaller clusters (cf. Fig.\ 8), so low-$N$ clusters spend a longer time in the higher
mass-loss phases and dissolve quicker. This steepens the scaling of the half-mass times.

The largest single-mass cluster studied dissolves in about 3.5 half-mass relaxation times. If the 
lifetimes of single-mass clusters continue to scale with $t_{rh}^{3/4}$, they would fall below one
relaxation time for clusters containing more than $N = 2.5 \; 10^6$ stars, which seems to be rather
unlikely. Several reasons could cause a change in the scaling before this point is reached:
First, we assume an evolution through equilibrium distributions. This assumption will certainly be
violated, since, if $t_{Diss} < t_{rh}$, clusters dissolve before any equilibrium can
be established. Second our assumption that the number of potential escapers at $\hat{E} = 0$ is proportional
to the number of cluster stars might be violated if escape becomes very efficient. Heggie (2000) constructed
a cluster in which the number of cluster stars is a function of $\hat{E}$ and $t$. Solving eq.\ 12, he could show
that the dissolution time scales with $t_{rh}$ if $N \rightarrow \infty$. However, this scaling is reached only
for particle numbers beyond the globular cluster regime.
 
Summarising, it is not clear whether the lifetimes still
scale with $t_{rh}^{3/4}$ if $N$ becomes much larger than $10^6$. Their scaling up to this point might
however be described by such a scaling law.
A similar value is found for the multi-mass clusters of the Collaborative Experiment.
The $t_{rh}^{3/4}$ scaling might therefore describe the scaling of the lifetimes for most of the
globular cluster regime. 
\begin{figure}
\epsfxsize=8.0cm
\begin{center}
\epsffile{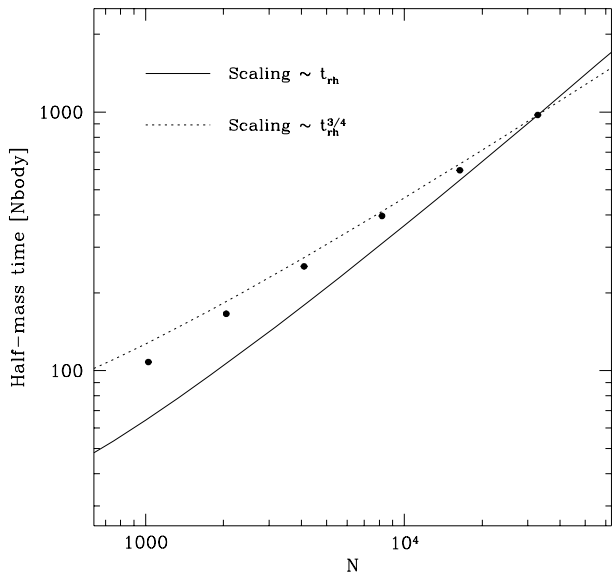}
\end{center}
\caption{Comparison of the predicted $t_{rh}^{3/4}$ scaling with the dissolution times of 
  multi-mass clusters from the Collaborative Experiment. Like in the case of single-mass  
   clusters, a $t_{rh}^{3/4}$ scaling law gives a fairly good fit to the half-mass times. 
     Multi-mass clusters show a stronger increase in their half-mass times than
      single-mass clusters due to their earlier core-collapse.}  
\end{figure}
 
\section{Conclusions}

The evolution of three different kinds of clusters was studied. It was found that the lifetime
scales with the relaxation time only if potential escapers are immediately removed. 
Otherwise, the lifetime increases more slowly with the particle number than the relaxation time. 
The reason for this discrepancy is that for radial cut-off clusters and for clusters in a 
tidal field, there is a difference in time between the moment 
when stars are scattered above the energy necessary for escape and the moment when they 
actually leave the cluster. During this time, potential escapers can be scattered back to 
energies below the critical one and remain bound. This backscattering of potential escapers 
causes a deviation from a scaling with the relaxation time.

For clusters limited by a radial cut-off, we expect this deviation to vanish for large 
enough $N$, since the time needed for escape increases only slowly with the particle number.
Since the relaxation time increases almost linear with $N$, it becomes very large compared 
to the escape time and all stars scattered above the critical energy leave the cluster,
causing the lifetime to scale with the relaxation time.       

Clusters in a steady tidal field show a larger discrepancy than the radial cut-off clusters.
This is due to the fact that in a tidal field the escape times depend on the energies 
of the stars and are large for stars with energies only slightly above the critical
one. Hence there are always stars that have escape times comparable to their 
relaxation times and many of them are scattered back to lower energies 
before they can escape.  

If we utilise the result of Fukushige \& Heggie \cite{fh},
namely that the escape time drops with the energy difference $E_{Star}-E_{Crit}$ to the 
second power, we expect that the 
lifetime scales with $t_{rh}^{3/4}$. Such a dependance gives a good fit to 
the half-mass times of the single-mass clusters studied here and the
multi-mass clusters of the Collaborative Experiment (Heggie et al.\ 1998).

We expect that there will be a transition for very large $N$ beyond which 
the lifetime scales with the relaxation time. This transition might
only play a role for the very largest globular clusters. Other processes, like         
for example the initial evolution until an 
equilibrium distribution of potential escapers is established and the core-collapse of 
the clusters influence the scaling of the lifetimes as well. 

Since the lifetimes increase more slowly with the number of cluster stars than the relaxation 
time, globular clusters will have shorter lifetimes than expected hitherto. More globular 
clusters might have been destroyed since the time of their formation and the remaining 
ones have suffered more from dynamical evolution. 

\section*{Acknowledgements}
I thank Douglas Heggie for many valuable comments and suggestions related to this 
work. I'm also grateful to Toshio Fukushige for the idea to Figure 3, Rainer Spurzem
for his help with the NBODY6++ code, and an anonymous referee for his suggestions
which improved the presentation of the paper.
It is a pleasure to acknowledge the support of 
the European Commission through TMR grant number ERB FMGE CT950051 (the TRACS  
Programme at EPCC). The parallel calculations were performed on the
CRAY T3E's of HLRZ J\"ulich and HLRS Stuttgart.
H.B.\ is supported by PPARC under grant 1998/00044.

\bsp
\label{lastpage}

\end{document}